\documentclass[reprint,prl,aps]{revtex4-2}

\usepackage[T1]{fontenc}
\usepackage[utf8]{inputenc}
\usepackage{color}
\usepackage{units}
\usepackage{amssymb}
\usepackage{graphicx}
\usepackage{esint}
\usepackage{bm}
\usepackage{natbib}
\usepackage[breaklinks=true,colorlinks=true,urlcolor=blue,
citecolor=blue,linkcolor=blue,bookmarks=false]{hyperref}
\usepackage{setspace}
\usepackage{ulem}
\usepackage{pdfpages}

\setcitestyle{journalcolor= blue}

\usepackage{babel}
\newcommand{\rxx}{\ensuremath{R_\mathrm{xx}}}
\newcommand{\ryx}{\ensuremath{R_\mathrm{xy}}}
\newcommand{\rh}{\ensuremath{\rho_{xy}}}
\newcommand{\rl}{\ensuremath{\rho_{xx}}}
\newcommand{\bperp}{\ensuremath{B_\perp}}
\newcommand{\bpar}{\ensuremath{B_\parallel}}
\newcommand{\mb}{moir\'e band}
\usepackage{babel}

\makeatletter
\AtBeginDocument{\let\LS@rot\@undefined}
\makeatother

\begin{document}
%%%%%%%%%%%%%%%%%%%%%%%%%%%% TITLE AND AUTHOR LIST %%%%%%%%%%%%%%%%%%%%%%%%%
\title{Spontaneous time-reversal symmetry breaking in twisted double bilayer graphene}
	
\author{Manabendra Kuiri$^{1}$}
\email{koolmanab@gmail.com}
\author{Christopher Coleman$^{1}$}
\author{Zhenxiang Gao$^{1}$}
\author{Aswin Vishnuradhan$^{1}$}
\author{Kenji Watanabe$^{2}$}
\author{Takashi Taniguchi$^{3}$}
\author{Jihang Zhu$^{4}$}
\author{Allan H. MacDonald$^{4}$}
\author{Joshua Folk$^{1}$}
\email{jfolk@physics.ubc.ca}
\affiliation{$^{1}$ Department of Physics and Astronomy \& Stewart Blusson Quantum Matter Institute,  University of British Columbia, Vancouver BC, Canada V6T 1Z4}
\affiliation{$^{2}$Research Center for Functional Materials, National Institute for Materials Science, Namiki 1-1, Tsukuba, Ibaraki 305-0044, Japan}
\affiliation{$^{3}$International Center for Materials Nanoarchitectonics,
National Institute for Materials Science, Namiki 1-1, Tsukuba, Ibaraki 305-0044, Japan}
\affiliation{$^{4}$ Physics Department, University of Texas at Austin, Austin TX USA 78712}

\date{\today}
	
\begin{abstract}
\textbf{Twisted double bilayer graphene (tDBG) comprises two Bernal-stacked bilayer graphene sheets with a twist between them. Gate voltages applied to top and back gates of a tDBG device tune both the flatness and topology of the electronic bands, enabling an unusual level of experimental control.
Broken spin/valley symmetry metallic states have been observed in tDBG devices with twist angles $\sim $ 1.2-1.3$^\circ$, but the topologies and order parameters of these states have remained unclear.   
We report the observation of an anomalous Hall effect in the correlated metal state of tDBG, with hysteresis loops spanning 100s of mT in out-of-plane magnetic field ($B_{\perp}$) that demonstrate spontaneously broken time-reversal symmetry. The \bperp~hysteresis persists for in-plane fields up to several Tesla, suggesting valley (orbital) ferromagnetism.  At the same time, the resistivity is strongly affected by even mT-scale values of in-plane magnetic field, pointing to spin-valley coupling or to a direct orbital coupling between in-plane field and the valley degree of freedom.}

\end{abstract}

\maketitle
The interplay between band topology and Coulomb interactions has emerged as a frontier in the study of two-dimensional (2D) materials with flat electronic bands\cite{cao2018correlated,cao2018unconventional, balents2020superconductivity,andrei2020graphene,PhysRevB.99.075127}.
Because they have native Dirac points close to the Fermi level, which provide a resource for band topology,
graphene-based van der Waals (vdW) heterostructures offer a flexible platform from which to investigate this interplay: flat bands may be readily engineered by clever heterostructure design\cite{bistritzer2011moire,cao2018correlated,cao2018unconventional} and tuned using experimental knobs such as magnetic field or gate voltage\cite{zhang2005experimental, PhysRevLett.122.016401}; the topology of the resulting bands is also readily controlled, whether by choice of heterostructure stacking or by tuning applied gate voltages\cite{PhysRevB.99.235406,lee2019theory}. Coulomb interactions 
frequently lead to spontaneous symmetry breaking  yielding exotic electronic phases such as fractional Chern insulators\cite{xie2021fractional} or unconventional superconductivity\cite{cao2018unconventional,oh2021evidence}.

The physical phenomenology that results from broken symmetry phases depends on the topology of the underlying electronic bands\cite{du2021engineering,PhysRevB.99.075127}. For example,  the anomalous Hall effect (AHE) is a striking experimental signature that emerges when spontaneous spin or valley polarization breaks time reversal symmetry in bands with finite Berry curvature\cite{RevModPhys.82.1539}. 
AHE reflecting orbital ferromagnetism  has been reported in multi-layer vdW heterostructures such as twisted bilayer graphene aligned with hexagonal boron nitride ($h$BN)\cite{sharpe2019emergent,serlin2019intrinsic}, and in naturally-occurring structures like Bernal-stacked (AB) bilayer graphene\cite{geisenhof2021quantum}.

%%%%%%%%%%%%%%%%%%%%% FIGURE 1 %%%%%%%%%%%%%%%%%%%%%%%%%%%%%%%%%%%%
\begin{figure}
	\begin{center}
		\includegraphics [width=1\linewidth]{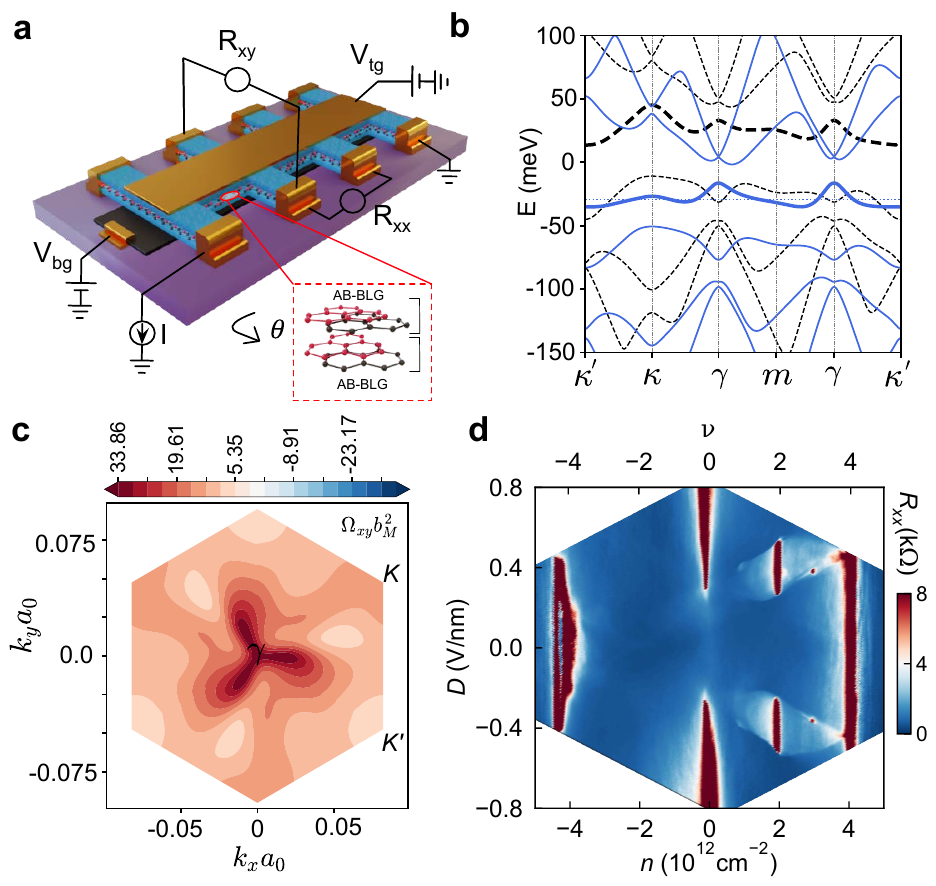}
	\end{center}
	\caption{{\bf{Twisted double bilayer graphene}} 
		(a) Schematic of a tDBG device consisting of two Bernal (AB) bilayer graphene stacked with a twist angle $\theta \approx1.3^{\circ}$. The stack is encapsulated with top and bottom hexagonal boron nitride ($h$BN) layers, with a graphite bottom gate ($V_{bg}$) and a metal topgate ($V_{tg}$) to independently tune the density ($n$) and displacement field ($D$).  (b) Calculated \mb~ dispersion for tDBG with twist angle $\theta=1.3^{\circ}$. The solid (dashed) line corresponds to Hartree–Fock (single particle) calculations. (c) Berry curvature ($\bf{\Omega}$) calculated for the conduction band.  (d) Four terminal resistance, $R_{xx}$ as a function of carrier density ($n$) and vertical displacement field ($D$) at $T$=20mK, and $B$=0 for device D1. The top axis shows the band filling, $\nu$. }
	\label{fig1}
\end{figure}
%%%%%%%%%%%%%%%%%%%%%%%%%%%%%%%%%%%%%%%%%%%%%%%%%%%%%%%%%%%%%%%%%%%%%

Twisted double bilayer graphene (tDBG)---two AB-stacked bilayer graphene sheets misaligned by twist angle $\theta$---is a uniquely tunable system in which band topology, correlations and broken symmetry phases can be  manipulated using top- and back-gate voltages to control the doping ($n$) and vertical displacement field ($D$) (Fig.~\ref{fig1}a)\cite{lee2019theory,PhysRevB.99.235406,PhysRevB.99.235417}.  When $\theta\sim 1.3^\circ$,
%the moiré conduction miniband
the moir\'e-modified conduction band (hereafter referred to as the \mb) can be tuned using $D$ to be nearly flat, separated from a dispersive band at higher energy and from the valence band below (Fig.~\ref{fig1}b)\cite{PhysRevLett.123.197702}. Theoretical predictions\cite{lee2019theory,PhysRevB.99.235406,PhysRevX.9.031021} and experimental data\cite{PhysRevLett.123.197702,shen2020correlated,liu2020tunable,cao2020tunable,he2021symmetry,wang2022bulk} indicate that, for this twist angle, the \mb~is 
topologically non-trivial and that interactions lead to broken spin and valley symmetries, similar to the symmetry breaking observed in quantum Hall ferromagnetism\cite{PhysRevLett.96.256602,young2012spin}.

The precise way in which spin and valley symmetries are broken in tDBG remains an open problem\cite{PhysRevLett.123.197702,shen2020correlated,he2021symmetry,lee2019theory}.  Valley-polarized, spin-valley-locked, spin-polarized, and intervalley coherent states have all been considered\cite{lee2019theory}.  The consistent observation of an insulating state when the \mb~is half filled is a valuable clue.   
The resistance of this state increases with in-plane field, \bpar, suggesting\cite{shen2020correlated,liu2020tunable} that the metallic states on either side may be spin polarized\cite{liu2020tunable,he2021symmetry} but leaving the question of possible valley order unaddressed\cite{PhysRevLett.123.197702,he2021symmetry,liu2020anomalous}.  
In particular, an AHE of the kind reported in other graphene systems has not been reported in tDBG with small $\theta$.

Here, we report the observation of a strong AHE in AB-AB stacked tDBG  ($\theta=1.3^{\circ}$), with longitudinal (\rxx) and Hall resistance ($R_{xy}$) that are sharply hysteretic under out-of-plane magnetic field (\bperp).  The data offer the first demonstration of orbital (valley) time reversal symmetry breaking in AB-AB tDBG.
Orbital order modifies the resistivity via strong Berry curvature, confirming predictions for tDBG near the top of the \mb~  (Fig.~\ref{fig1}c). Extreme asymmetry between in- and out-of-plane magnetic field dependence in the AHE signatures reinforces
the orbital (valley) character of the ferromagnetism\cite{tschirhart2021imaging,sharpe2021evidence}.   At the same time, both longitudinal and transverse resistivity are sensitive to unusually small in-plane magnetic fields, an effect that is not yet understood.

Devices were fabricated using established techniques\cite{kim2016van,cao2018correlated} (see Methods), and patterned into Hall bar geometries to measure $R_{xx}$ and $R_{xy}$. Fig.~\ref{fig1}a shows a schematic of the device architecture and measurement scheme.  Multiple voltage probe pairs were measured in each device, with similar behaviour across most pairs (see SI).  Mixing between $R_{xx}$ and $R_{xy}$ have been
minimized in the data as presented by reporting the field symmetrized longitudinal resistances,
henceforth labelled $\rho_{xx}$ and
$\rho_{xy}$, respectively (see Methods). Reported values of $\bperp$ have been adjusted to reflect flux trapping in the superconducting magnet (see SI).
Similar behaviour is seen in two devices (D1 and D2), with twist angles $\theta\approx1.31^{\circ}$ and $\theta \approx1.34^{\circ}$ respectively; these were the only two devices with twist angle near 1.3$^\circ$ that were measured.

A typical resistivity map over top- and back-gate is shown in Fig.~\ref{fig1}d, plotted with respect to $n$ and $D$, see Methods.  Insulating stripes at $n=0$ reflect the separation of conduction from valence band by finite $D$, while the insulating stripe around $n=4\times10^{12}$ cm$^{-2}$ reflects full filling of the first conduction band. Given the 4-fold degeneracy of the band (spin and valley), full filling is achieved at four electrons per moir\'e cell ($\nu=4$), allowing 
the twist angle to be calculated and the relation between $n$ and $\nu$ determined (see Methods).  Near $D\sim\pm 0.4$ V/nm, numerical calculations indicate a \mb~that is nearly flat and isolated both from lower and upper bands. Fig.~\ref{fig1}b shows single-particle and self-consistent Hartree calculations of the \mb~structure for the relevant parameters, illustrating the flatness of the band especially in the self-consistent calculation that enhances the role of interactions\cite{PhysRevLett.123.197702,science.abc3534}.

Setting $D$ near $\pm 0.4$ V/nm in the device, a strong insulating state at $\nu=2$ is clearly visible, consistent with previous reports\cite{shen2020correlated,liu2020tunable,he2021symmetry},  with surrounding 'halo' regions of higher resistivity sharply separated from a lower-resistance background.  
The high quality of these devices is demonstrated by the 
the strong insulating state at $\nu=3$ even at $B=0$; this signature is known to emerge only over a narrow range of twist angle $\theta\sim 1.3^{\circ}$\cite{he2021symmetry} where the correlations are maximum.

%%%%%%%%%%%%%%%%%%%%% FIGURE 2 %%%%%%%%%%%%%%%%%%%%%%%%%%%%%%%%%%%%
\begin{figure*}
% 	\begin{center}
		\includegraphics [width=1\linewidth]{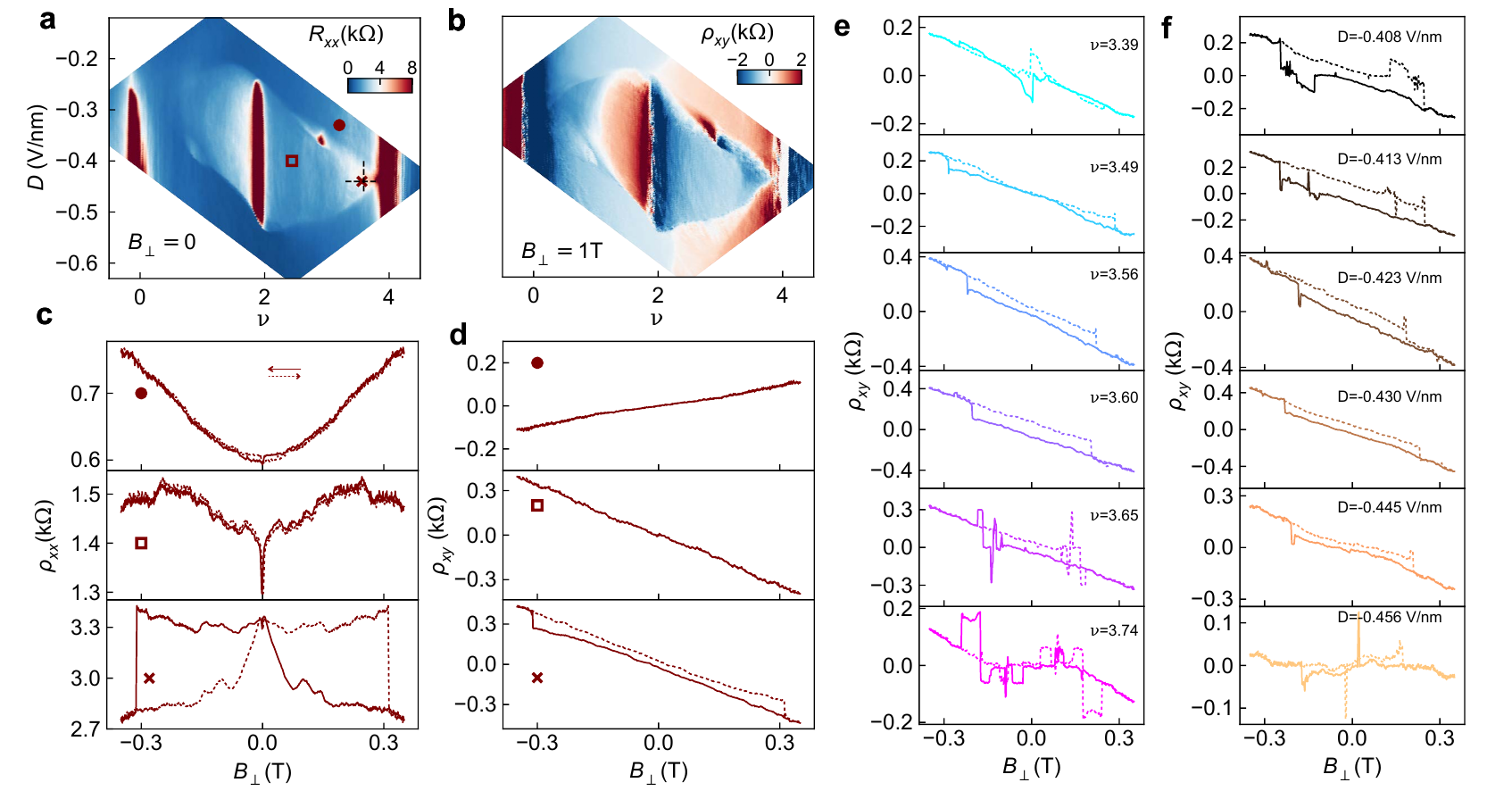}
% 	\end{center}
	\caption{ {\bf{Out-of-plane magnetoresistance.}} 
		(a) $R_{xx}$ as function of $\nu$ and $D$ showing the correlated insulating states at $\nu=2$ and $\nu=3$ surrounded by a halo region of higher resistance ($B$=0). (b) Anti-symmetrized Hall resistance,  $\rho_{xy}$ for $|B_\perp|=1$T. The Hall resistance changes sign at all integer fillings and also at the boundary of the halo region. (c) and (d) Magnetic field dependence of the symmetrized longitudinal resistance, $\rho_{xx}(B_\perp)$, and anti-symmetrized Hall resistance, $\rho_{xy}(B_\perp)$, for three values of \{$\nu,D$\} marked in (a).  $B_{\perp}$ is swept back and forth, shown with solid (positive to negative) and dashed (negative to positive) lines. (e) and (f) $\rho_{xy}(B_\perp)$ as $B_\perp$ is swept back and forth (solid/dashed) for (e) varying fillings $\nu=3.4\rightarrow 3.75$ at fixed $D=-0.43$V/nm, and (f) varying $D=-0.4\rightarrow =-0.46$ V/nm at fixed $\nu=3.6$. $T=20$ mK for all.}
	\label{fig2}
\end{figure*}

%%%%%%%%%%%%%%%%%%%%% FIGURE 3 %%%%%%%%%%%%%%%%%%%%%%%%%%%%%%%%%%%%
\begin{figure}
\begin{center}
\includegraphics [width=1\linewidth]{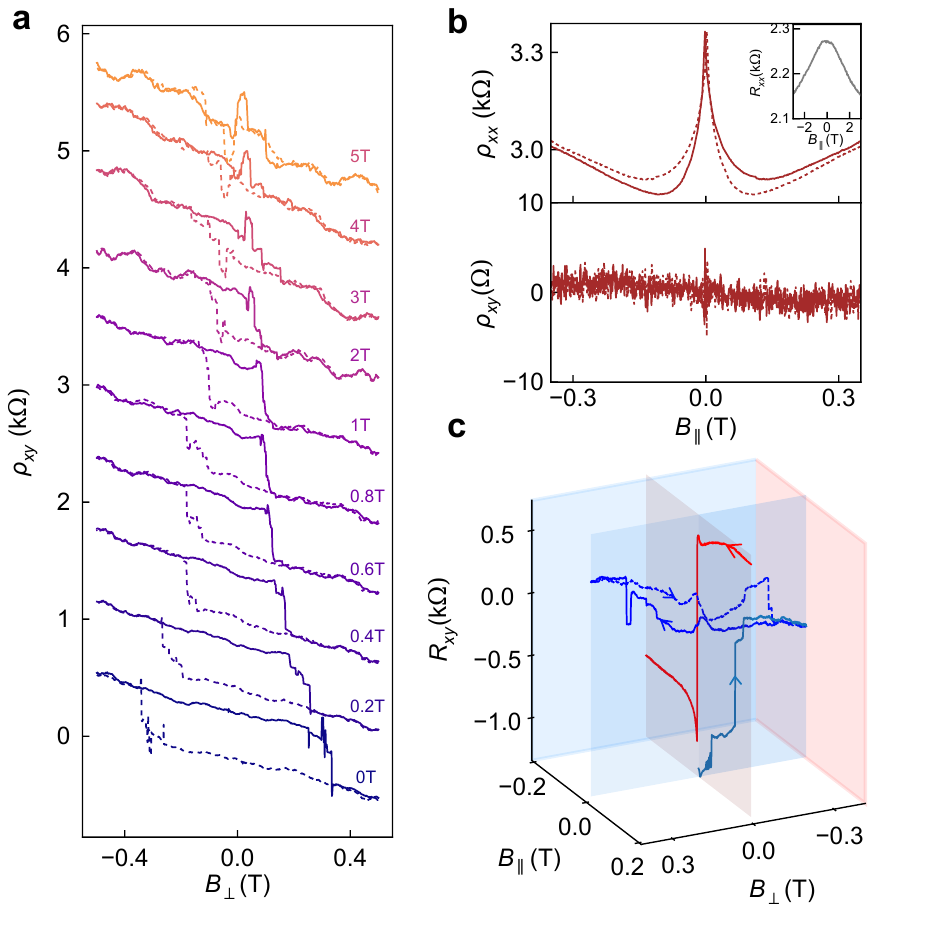}
\end{center}
\caption{{\bf{Anisotropic magnetoresistance.}}
	(a)  Out-of-plane  hysteresis $\rho_{xy}(\bperp)$ for fixed in-plane magnetic fields ($B_{\parallel}$) up to 5 T.  (D2, $\nu=3.57$,  $D=-0.43$V/nm) The curves are offset by 600$\Omega$ for clarity. (b and c) Two examples of extreme in-plane magnetoresistance: (b) For some gate voltage settings, a sharp non-hysteretic peak was observed in $\rxx(\bpar)$, while \ryx~showed minimal magnetoresistance (D1, $\nu=3.43$, $D=-0.42$V/nm). Slight differences between up-sweep and down-sweep represent mild heating crossing $\bpar=0$.  Inset highlights sharpness of resistance peak. (c) For other gate voltage settings, sweeping \bpar~induced a large switch in \ryx~(red trace, $\bpar=0.2\rightarrow -0.2$ T while $\bperp=0$).  Returning to $\bpar=0$, \ryx~remained large and negative, then sweeping $\bperp=0\rightarrow -0.3$T (grey trace) brought \ryx~back closer to zero, and subsequent \bperp~sweeps (blue traces) were again hysteretic as in panel (a) (D2, $\nu=3.57$, $D=-0.43$V/nm). $T=20$ mK for all.}
	\label{fig3}
\end{figure}

Figure~\ref{fig2}a offers a higher resolution map of $R_{xx}$ across the halo region at negative $D$, with insulating states at $\nu=2$ and $\nu=3$ clearly visible.
More insight into the broken-symmetry metallic states within the halo may be obtained from the Hall resistance, shown for $\bperp=\pm1T$ in Fig.~\ref{fig2}b. \rh~changes sign at $\nu=2$ across the entire halo, and also at $\nu=3$ across a narrow region at the low-$D$ edge of the halo.  This behaviour implies that the 4-fold band 
degeneracy is broken throughout the halo region, and fully lifted 
for a narrow range of $D$.

The most telling data comes from the magnetoresistance of these correlated metallic states.  Figs.~\ref{fig2}c and \ref{fig2}d show \rl~and \rh~ as  \bperp~ is swept from 350 mT to -350 mT and back, comparing data for three $\{\nu,D\}$ pairs, indicated by markers in Fig.~\ref{fig2}a.  These datasets are chosen to illustrate three characteristic behaviours observed within the $\{\nu,D\}$ map. Outside the halo region ($\bullet$), longitudinal magnetoresistance is weak and the data are independent of sweep direction, consistent with the behaviour of non-interacting metals.  Within the halo but far from $\nu=4$ ($\square$), \rl~shows an additional strong but narrow positive magnetoresistance, but \rh~is again nearly linear and neither \rl(\bperp) nor \rh(\bperp) depend on sweep direction.  Within the halo and near $\nu=4$ ($\times$), both \rh~and \rl~are strongly hysteretic, a clear indication of spontaneously broken time reversal symmetry giving rise to ferromagnetism with coercive fields near 0.3 T (bottom panels in Fig.~\ref{fig2}c,d). AHE was also observed for the halo at positive $D$, though the hysteresis was less pronounced (see SI).

%%%%%%%%%%%%%%%%%%%%% FIGURE 4 %%%%%%%%%%%%%%%%%%%%%%%%%%%%%%%%%%%%
\begin{figure}
	\begin{center}
		\includegraphics [width=1\linewidth]{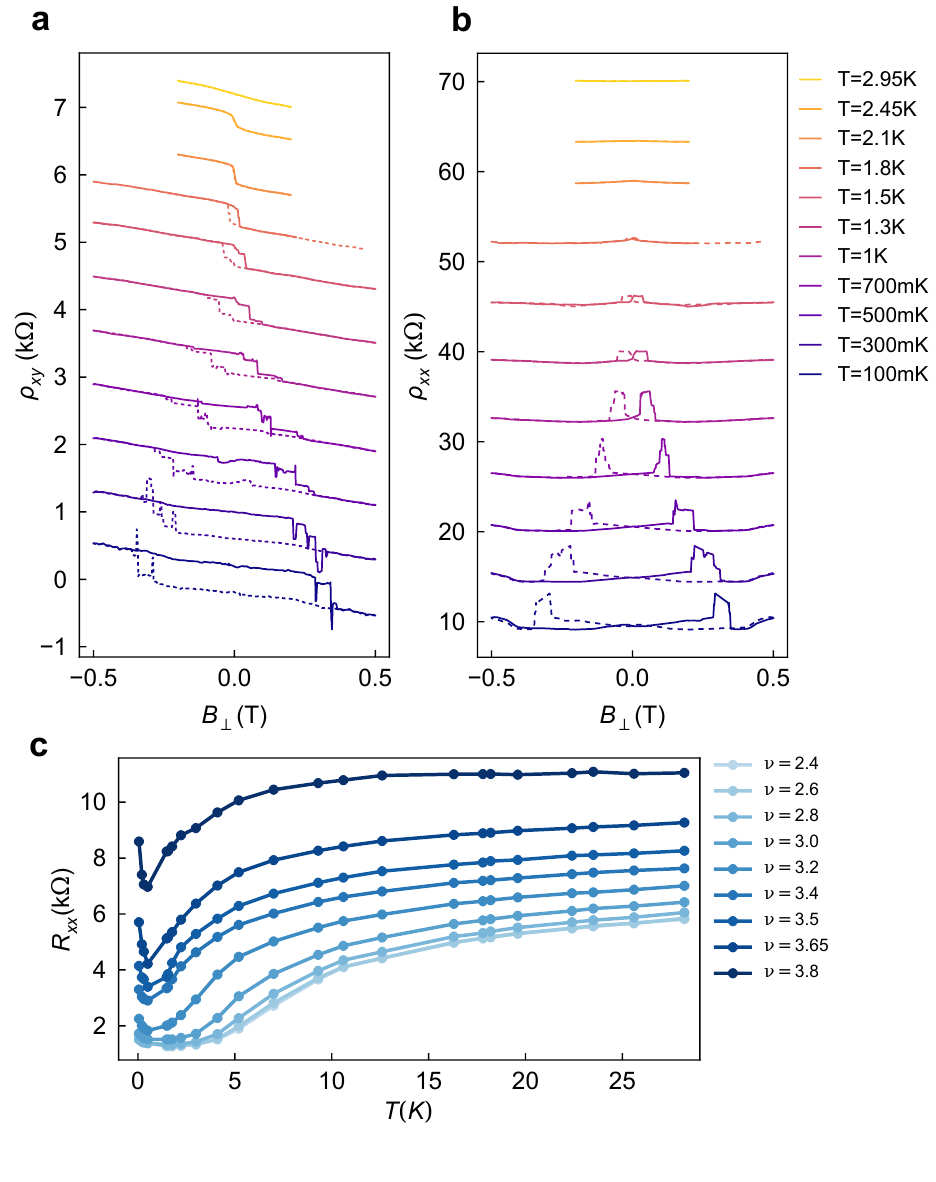}
	\end{center}
	\caption{ {\bf{Temperature dependence of  AHE.}} 
		(a) Temperature dependence of hysteresis loop  $\rho_{xy}(\bperp)$  (D2, $\nu=$3.57, $D=-$0.43 V/nm). Curves are offset by 0.8k$\Omega$ for clarity.  (b) Temperature dependence of $\rho_{xx}(\bperp)$; curves offset by 6$k\Omega$ for clarity. (c) Temperature dependence $R_{xx}(T)$ at $B$=0 in the halo region for D1, $D=-$0.43 V/nm, for $\nu$ in the range $\nu=$2.4-3.8.}
	\label{fig4}
\end{figure}

%%%%%%%%%%%%%%%%%%%%% FIGURE 3 %%%%%%%%%%%%%%%%%%%%%%%%%%%%%%%%%%%%

The range of $\nu$ over which ferromagnetic AHE signals appear is illustrated in Fig.~\ref{fig2}e, showing data along the $D=-0.43$ V/nm line in Fig.~\ref{fig2}a.  Close to the single-particle insulator ($\nu>3.75$) multiple jumps are seen in up-and-down magnetic field sweeps, indicating multi-domain switching behaviour; this behaviour continues deep into the insulating state.   Throughout the range $3.5\lesssim\nu\lesssim3.7$, the hysteresis loop was wide and clean with nearly constant coercive field, then below $\nu\sim 3.4$ both the width and height of the hysteresis loop collapsed, with no AHE seen for $\nu<3.3$.  The reduced width of the loops indicates lower coercive fields for $\nu\lesssim 3.4$. 
The reduced height may result from reduced valley polarization or smaller Berry curvature: numerical calculations indicate that the strongest Berry curvature is near $k_x=k_y=0$, corresponding to the highest energy (last-filled) states in the \mb  (Fig.~\ref{fig1}c).

Interestingly, the hysteresis loops were nearly independent of $D$ within the halo.  The data in Fig.~\ref{fig2}f represent the evolution of the AHE at fixed $\nu=3.6$, varying the displacement field from  $D=-0.4$~V/nm to $D=-0.46$~V/nm (Fig.~\ref{fig2}f).  Although the AHE disappeared abruptly outside of the halo, across the halo there was little change with $D$.  This uniformity contrasts with the very narrow range of $D$ over which the $\nu=3$ insulating state is observed.

Having established a ferromagnetic AHE close to full filling of the \mb, we turn to the question of whether spin or valley symmetry breaking is responsible for the observed effect.  Previous reports of spin polarization in the $\nu=2$ insulator and presumably in the surrounding metallic states might suggest spin ferromagnetism as a possible source for the AHE, but the weak spin-orbit interaction\cite{PhysRevLett.122.046403} of graphene makes this explanation less likely.  Experimental input into this question is obtained from a comparison of in-plane and out-of-plane magnetoresistance: whereas the valley degree of freedom couples primarily to out-of-plane field, spins are expected to couple to total magnetic field with $g=\sim2$.  

Figure~\ref{fig3} explores the effect of \bpar~in more detail.  A clear illustration of the anisotropic nature of the AHE comes from Fig.~\ref{fig3}a, where \bperp~hysteresis loops ($\pm450$ mT) are shown for increasing fixed \bpar.  A particularly robust AHE in device D2 exhibits a small hysteresis loop in \bperp, with coercive field around $\sim$50 mT, even when \bpar~ is held at 5T.  (Equivalent data for Device D1 show \bperp~ hysteresis persisting above $\bpar\sim 1$ T, see SI.)  These behaviours are reminiscent of the angle-dependent hysteretic AHE observed in twisted bilayer graphene aligned to $h$BN\cite{sharpe2021evidence}, where the impact of in-plane component of $B$ was similarly weak up to fields of several T.  In that system the AHE signal was attributed to valley polarization that is weakly sensitive to \bpar.

In our measurements of $t$DBG, there is a qualitative distinction between phenomena driven by \bperp~and \bpar~but the influence of \bpar~is nevertheless very strong---much stronger for small \bpar~than in Ref.~\onlinecite{sharpe2021evidence}.  Figs.~\ref{fig3}b,c illustrate two types of extreme \bpar~magnetoresistance, both of which were observed in both devices. Throughout most of the halo region an anomalously sharp negative \bpar-magnetoresistance was observed (Fig.~\ref{fig3}b), with a rounding around zero field on the $\bpar\sim 1$ mT scale that is orders of magnitude smaller than what might be expected when comparing $g\mu_B \bpar$ to $k_B T$ ($k_B T/g \mu_B \sim 100$ mT at 100 mK for $g$=2). This effect cannot be explained by experimental artifacts as it was absent for gate settings outside the halo region.

For some gate voltage settings within the AHE region, an even more surprising \bpar~dependence was observed (Fig.~\ref{fig3}c), with magnetic field sweeps through $\bpar=0$ leading to jumps in $R_{xy}$ up to several k$\Omega$, which could be reset by subsequent hysteresis loops in \bperp. This behaviour was observed in the AHE regions of devices D1 and D2, though it was significantly more prevalent in D2. The striking effects of small \bpar~ shown in Fig.~\ref{fig3} have not, to our knowledge, been reported before; they appear in both devices, and only inside the halo regions. 
These effects of \bpar~may be due to its coupling to valley order\cite{lee2019theory}.  Even though this coupling is small, proportional to the  thickness of the double bilayer, it competes with valley anisotropy which is poorly understood at present and likely also to be weak.
Alternately, \bpar~ could influence the AHE via spin-orbit couping as reported
recently in twisted (single) bilayer graphene with spin-orbit interaction enhanced by proximity to WSe$_2$\cite{lin2022spin}.  Although spin-orbit coupling was not intentionally enhanced in our devices, this possibility cannot be excluded.
A more complete study of the \bpar-induced effects in tDBG will be the subject of future work.

The AHE signatures seen in Fig.~\ref{fig2} persist up to 1.8 K in device D2 (similar in D1), illustrating the relatively large energy scales associated with valley symmetry breaking in this system.  Figs.~\ref{fig4}a and b show the temperature-induced collapse of the hysteresis loop via shrinking coercive field for \rxx~and \ryx~respectively. Early measurements of tDBG noted a step-like rise in \rxx~with temperature within the halo region, which has been attributed to temperature-induced collapse of broken symmetry states\cite{he2021symmetry}.  An analogous rise in \rxx~ around 7 K is observed in our measurements away from the AHE region ($\nu<3.2$, Fig.~\ref{fig4}c). The onset of AHE with increasing $\nu$ is correlated with a sharp decrease in the transition temperature, consistent with the Kelvin-scale collapse of the symmetry-broken state that gives rise to the hysteresis in Figs.~\ref{fig4}a,b.  The gradual shift of the transition temperature with increasing $\nu$ in Fig.~\ref{fig4}, instead of the appearance of a second symmetry breaking transition in the AHE region, supports the notion of a correlated low-$T$ ground state that with coupled spin and valley order\cite{10.1143/PTP.16.58,PhysRevB.74.092401}.\\

In conclusion, we have observed an AHE, signifying orbital magnetic order, in AB-AB stacked tDBG. 
The ferromagnetic state occurs only within the strongly interacting `halo' regions of the  ($\nu,D$) plane. 
Strong magnetic anisotropy suggests valley ferromagnetism, while in-plane field signatures indicate a complex interplay of broken spin/valley 
symmetries in the correlated metallic state of tDBG.\\

\noindent\textbf{METHODS}\\

\textbf{Device Fabrication:}

High quality AB-AB stacked tDBG devices were fabricated using the 'modified' tear and stack technique\cite{kim2017tunable,cao2018correlated}. At first, a large bilayer graphene (BLG) flake $\sim 70\mu$m was exfoliated on a Si/SiO$_2$ substrate with 285nm oxide. The BLG flake was mechanically precut into two pieces using a sharp $\sim 1\mu$m diameter Tungsten dissecting needle which was attached to the micromanipulator in our transfer setup. Using a stamp made of PolyBisphenol carbonate (PC) on polydimethylsiloxane (PDMS), we picked up the top $h$BN layer, at $T$=100C$^{\circ}$. Next, BLG flake was picked up at $T$=30C$^{\circ}$, and the stage was rotated to 1.36$^{\circ}$, followed by picking up the second BLG at $T$=30C$^{\circ}$. Then the bottom $h$BN layer was picked up at $T$=100$^{\circ}$C, followed by graphite at $T$=110$^{\circ}$C. The sequence of stacking from top to bottom is $h$BN, tDBG, $h$BN and graphite. Finally, the entire stack was deposited on a Si/SiO$_2$ substrate at $T$=175$^{\circ}$C. Top gate was defined using electron beam lithography (EBL), followed by deposition of Cr/Au (5nm/50nm). Edge contacts were established by etching the stack in CHF$_3$:O$_2$ plasma, followed by deposition of Cr/Au (5nm/80nm) at a base pressure of $\sim 1\times 10^{-7}$mbar. Finally, another step of EBL and etching was done to pattern the device in a Hall bar geometry. The top and bottom $h$BN thickness was determined using atomic force microscopy.

\textbf{Transport measurements:}
The transport measurements were performed in four terminal configuration, using standard lock in (SRS830) techniques at $f\sim 13.3$Hz with a current bias $I=$1nA. The experiment was carried out in a Bluefors cryogen-free dilution refrigerators equipped with a 3-axis magnet enabling simultaneous control of in- and out-of-plane magnetic fields. Except where noted, measurements were carried out at the nominal base temperature of the refrigerator, with the mixing chanber at 10-15 mK, but heating due to magnetic field sweeps raised the sample temperature up to 50-200 mK depending on sweep conditions.

Twist angle, $\theta$, was calculated from the density corresponding to full filling of the moiré miniband $n(\nu= \pm 4)$ = 8$\theta^{2} / \sqrt{3} a^{2}$, where $a$ = 0.246 nm is the lattice constant of graphene. In our device, D1 $\nu=\pm 4$ corresponds $n=\pm ~4.05\times 10^{12}\text{cm}^{-12}$, which translates to a twist angle $\theta\approx 1.31^{\circ}$. The top-gate ($V_{tg}$) and bottom-gate ($V_{bg}$) voltages were used to independently control the density ($n$) and the displacement field ($D$). The charge density is given by $n=(C_{bg}V_{bg}+C_{tg}V_{tg})/e$ and the displacement field is given by $D=|C_{bg}V_{bg}-C_{tg}V_{tg}|/2\epsilon_0$, where $C_{bg}$ ($C_{tg}$) are the bottom(top) gate capacitance's per unit area, $e$ being the electronic charge and $\epsilon_0$ is the vacuum permittivity. We label the field symmetrized (anti-symmetrized) data for  $R_{xx}$ ($R_{xy}$) as $\rho_{xx}$ ($\rho_{xy}$), where $\rho_{xx}(B,\uparrow)=\frac{R_{xx} (B, \uparrow)+R_{xx} (B, \downarrow)}{2}\frac{W}{L}$, $\rho_{xx}(B,\downarrow)=\frac{R_{xx} (B, \downarrow)+R_{xx} (B, \uparrow)}{2}\frac{W}{L}$  and the anti-symmetrized Hall resistance $\rho_{xy}(B,\uparrow)=\frac{R_{xy} (B, \uparrow)-R_{xy} (B, \downarrow)}{2}$, $\rho_{xy}(B,\downarrow)=\frac{R_{xy} (B, \downarrow)-R_{xy} (B, \uparrow)}{2}$, using $\uparrow,\downarrow$ to indicate the magnetic field sweep direction.

\section{Acknowledgments}
We thank Leonid Levitov and Marcel Franz for fruitful
discussions and comments, and Ruiheng Su
for help with figure illustrations. MK acknowledges a postdoctoral research fellowship from
Stewart Blusson Quantum Matter Institute, UBC.  JF acknowledges funding from NSERC, CIFAR, SBQMI, CFI, and ERC Synergy funding for Project  No 941541.
AHM was supported by was supported by the U.S. Department of Energy, Office of Science, Basic Energy Sciences, under Awards No DE‐SC0019481 and DE-SC0022106.  K.W. and T.T. acknowledge support from JSPS KAKENHI (Grant Numbers 19H05790, 20H00354 and 21H05233).

\section{Author Contribution}
MK fabricated the devices, with help from AV; MK, CC, and ZG performed transport measurements and analysed the data; MK, JF, JZ and AM interpreted the data; MK and JF wrote the manuscript.  KW and TT provided the $h$BN crystals.

%\bibliography{ref_tdblg}


\onecolumngrid
\newpage
\thispagestyle{empty}
\mbox{}


\section*{Supplementary material}

# I. Device Fabrication

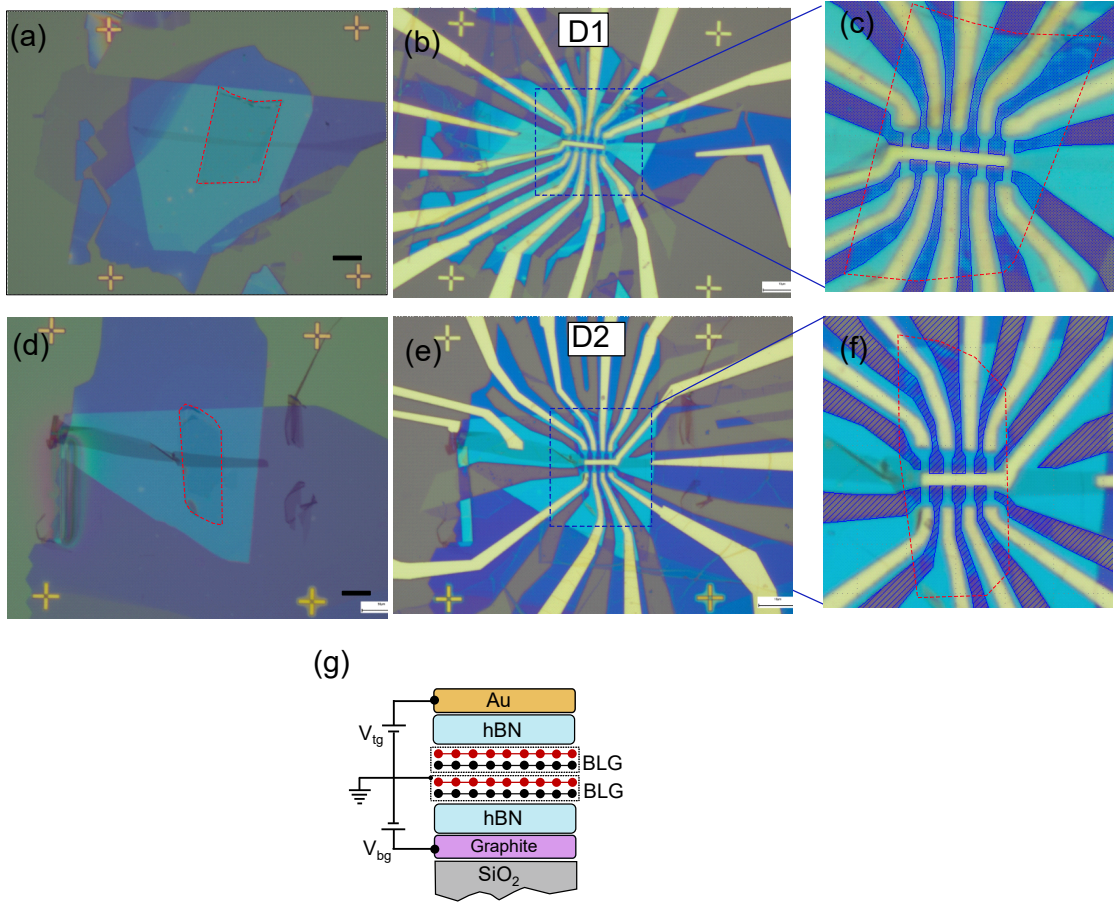

FIG. S1. **Device Fabrication.** (a) Optical image of the device D1 consisting of $h$BN/tDBG/$h$BN/graphite with a twist angle $\theta \approx 1.31°$ after stacking. tDBG boundary highlighted by red dashed line Scale bar is $7\mu$m. (b) Optical image of the device D1, after nanofabrication. (c) Zoomed-in optical image of the device D1, showing the contacts. The etched region has been shown in blue. (d) Optical image of second device D2 consisting of $h$BN/tDBG/$h$BN/graphite with $\theta \approx 1.34°$. Scale bar is $7\mu$m. (e) Optical image of the device D2, after nanofabrication. (f) Zoomed-in optical image of the device D2, showing the contacts. The etched region has been shown in blue. In both devices, the graphite layer serves as bottom gate, $V_{bg}$. The samples were not intentionally aligned to $h$BN, and accidental alignment can be ruled out because both devices (D1 and D2) show a similar behavior, and transport characteristics were the same for both positive and negative $D$.[S1]. (g) Cross section schematic of the devices.

## II. Sample uniformity and device characterization.

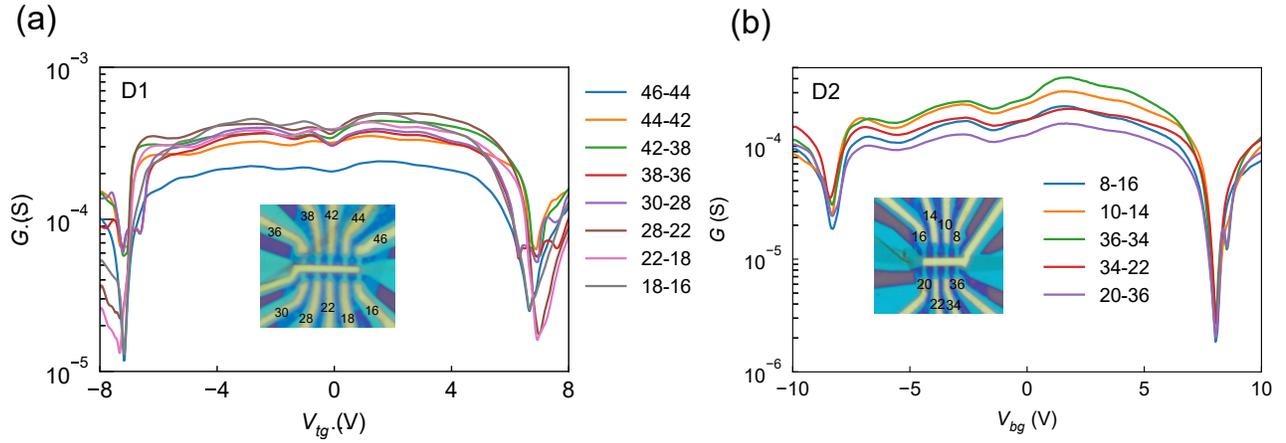

FIG. S2. **Sample uniformity and device characterization.** (a) Two wire conductance measurements between all neighbouring contact pairs at $T =4$K for device D1. The inset shows the optical image of the device with contact pairs labelled. Here, an ac excitation voltage of $V_{ac} = 100\mu$V was applied and the current was measured via the other contact, while all other contacts remained floating. The difference between the position of the moiré peak is used to estimate the local twist angle. The variation of $\delta V_{tg}(\delta n)$ for different contact pairs for D1 is $\delta n \sim 9 \times 10^{-10}$cm$^{-2}$, which translates to an angle inhomogeneity of $\delta\theta \sim 0.02°$ over a length of $12\mu$m. The slightly anomalous pair 46-44 was not used in this experiment. (b) Two wire conductance measurement for device D2 at $T =4$K. The twist angle inhomogeneity in this device, D2 is $\delta\theta \sim 0.01°$. The corresponding contact pairs are labelled in the inset.

## III. Four terminal measurements (D1).

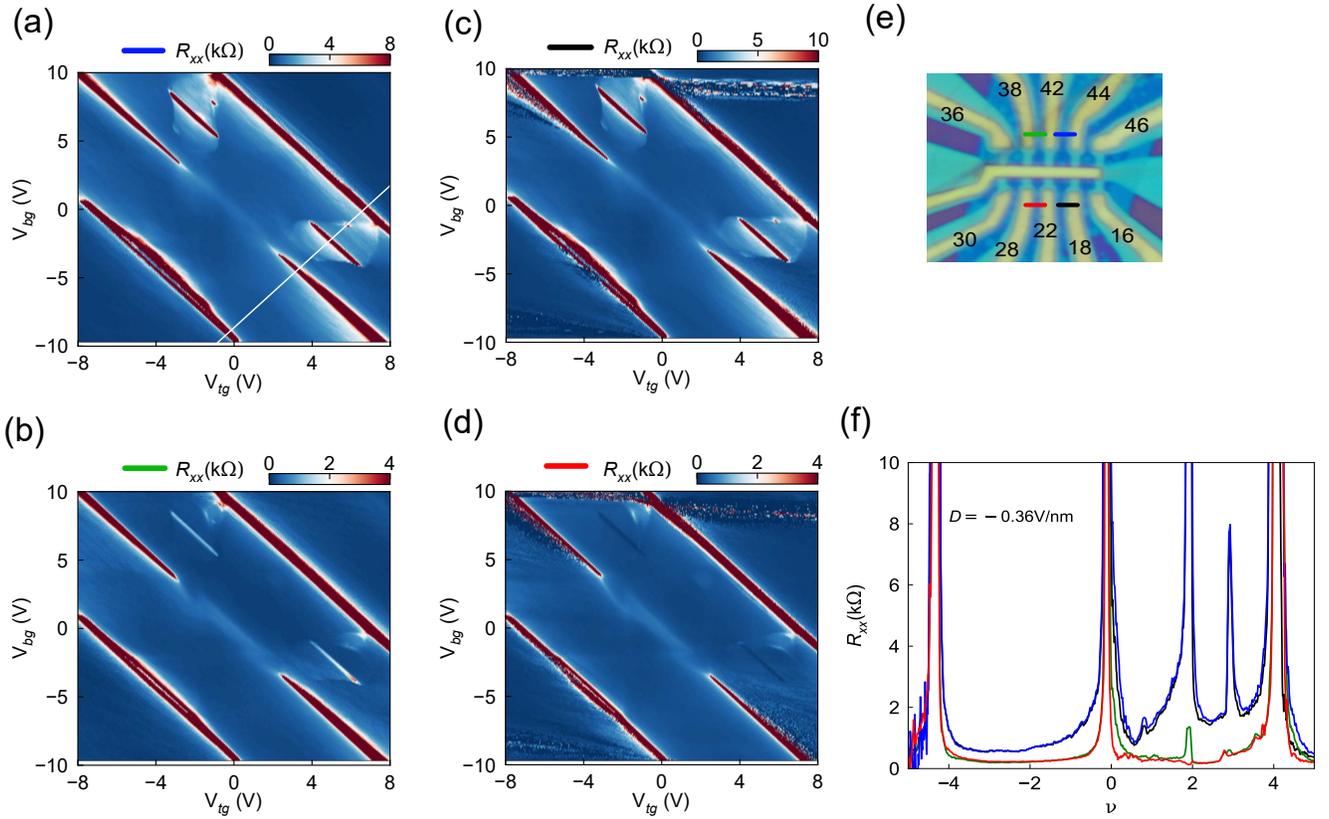

FIG. S3. **Comparison for different voltage probes: device 1 at** $B = 0$. (a)-(d) 2D colorplot of the measured Resistance as a function of topgate voltage ($V_{tg}$) and backgate voltage ($V_{bg}$) at $B = 0$, and $T=20$mK, for device D1 for several pair of contacts color coded in (e). The data in presented in Fig.1d of the manuscript, transformed to axes of $\nu$ and $D$, is from this figure panel (a). (f) Four terminal resistance as a function of filling factor, $\nu$ at $D = -0.36$ V/nm for several voltage probes, color coded in the inset. Here, current was injected in contact 16, and contact 30 was grounded. Most of the contact pairs show insulating states at filling $\nu = 0, 2, 3, \pm 4$, while the contact pair marked in red showed a suppression of the correlated insulating state at $\nu = 2$, with a resistance $\sim 220\Omega$. The white dashed line in (a) represents the line trace at $D = -0.36$ V/nm in the 2D map.

## IV. Four terminal measurements (D2).

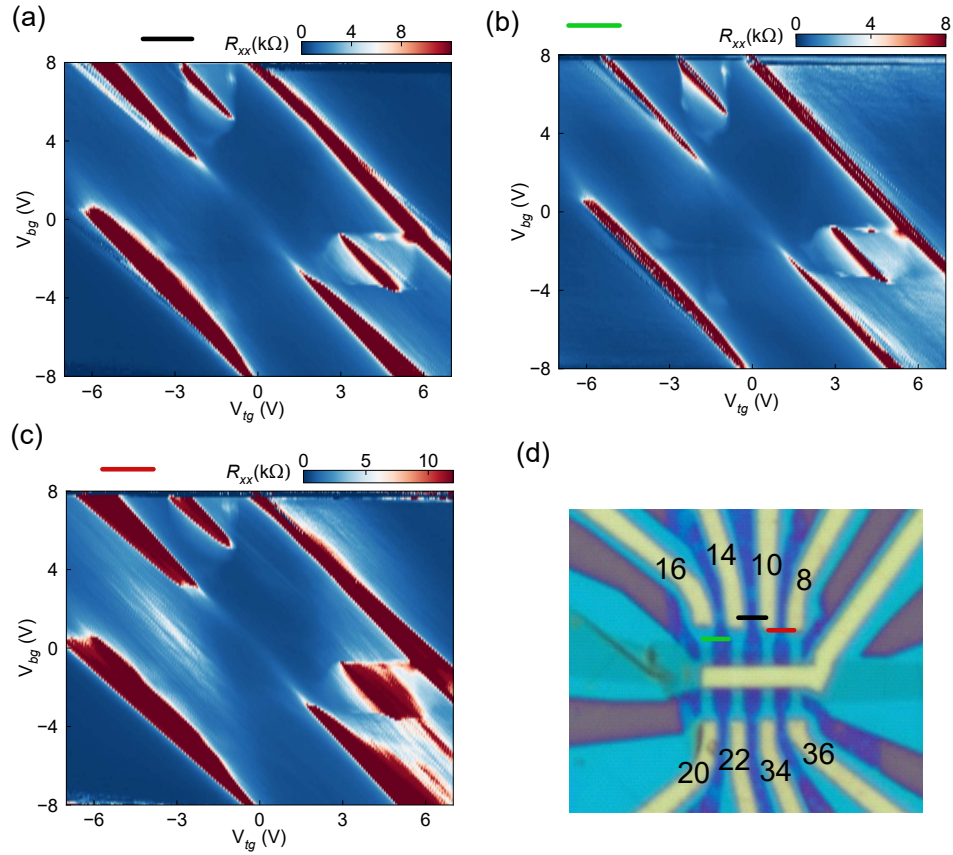

FIG. S4. **Comparison for different voltage probes: device D2 at $B = 0$.** (a)- (c) 2D colorplot of the measured resistance as a function of topgate voltage ($V_{tg}$) and backgate voltage ($V_{tg}$) at $B = 0$, and $T=20$mK, for several voltage pairs for device D2. The corresponding voltage pairs are color coded in (d).

## V. Hysteresis in different voltage probes: device D1

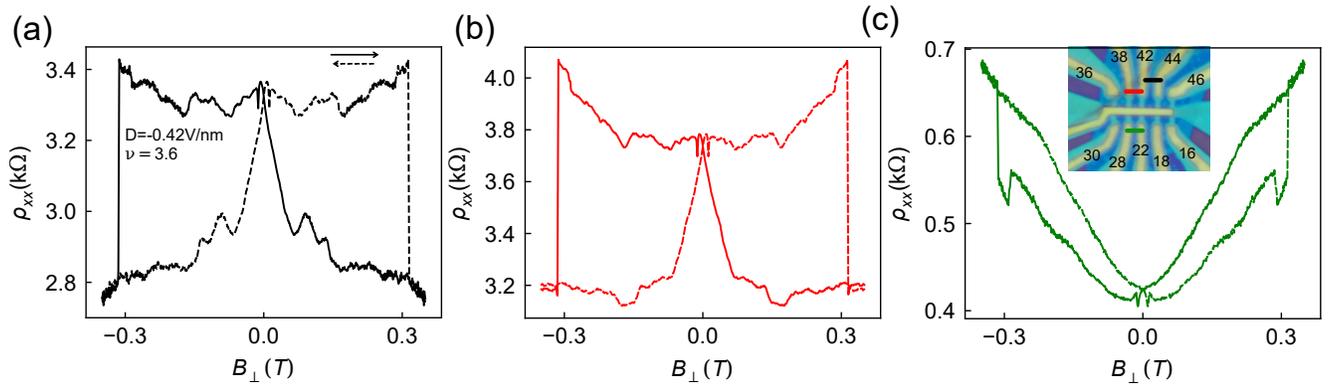

FIG. S5. **Comparison for different voltage probes: device D1.** (a) $\rho_{xx}$ as a function of out of plane magnetic field for $\nu = 3.6$, $D = -0.42$V/nm, where the magnetic field is swept back and forth for three voltage probes color coded in the inset of (c).

## VI. Hysteresis in different voltage probes: device D2

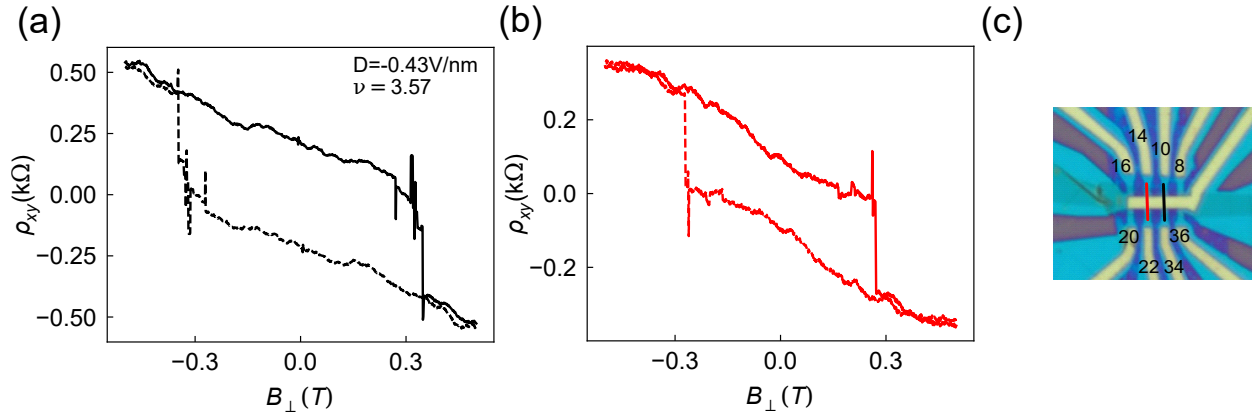

FIG. S6. **Comparison for different voltage probes: device D2.** (a) and (b) $\rho_{xy}$ as a function of out of plane magnetic field for $\nu = 3.57$, $D = -0.43$V/nm, where the magnetic field is swept back and forth, for the contact pairs color coded in (c).

## VII. Symmetrization and Anti-symmetrization of the measured longitudinal and Hall resistance, device D1

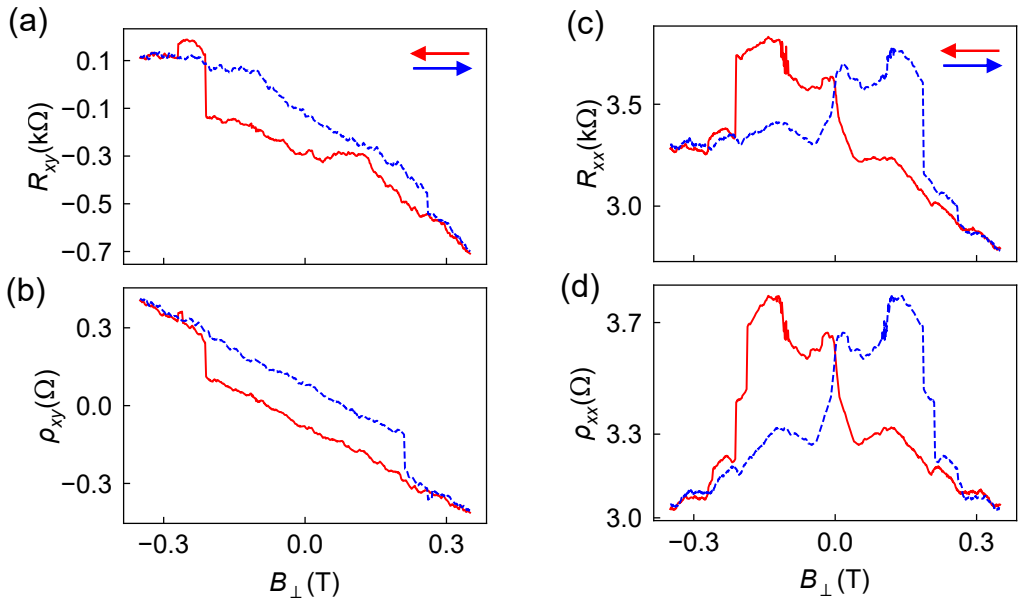

FIG. S7. **Symmetrization of $R_{xx}$ and Anti-symmetrization of $R_{xy}$.** (a) Measured Hall resistance $R_{xy}$ as a function of $B_\perp$, where the magnetic field is swept from $+B_\perp$ to $-B_\perp$ (red solid curve) and $-B_\perp$ to $+B_\perp$ (blue dashed curve). (b) Shows the corresponding anti-symmetrized Hall Resistance, $\rho_{xy}$. (c) Shows the measured longitudinal resistance $R_{xx}$. (d) Shows the symmetrized longitudinal resistance, $\rho_{xx}$. These data were taken at filling $\nu = 3.64$ and $D = -0.43$V/nm.

## VIII. Symmetrization and Anti-symmetrization of the measured longitudinal and Hall resistance, device D2

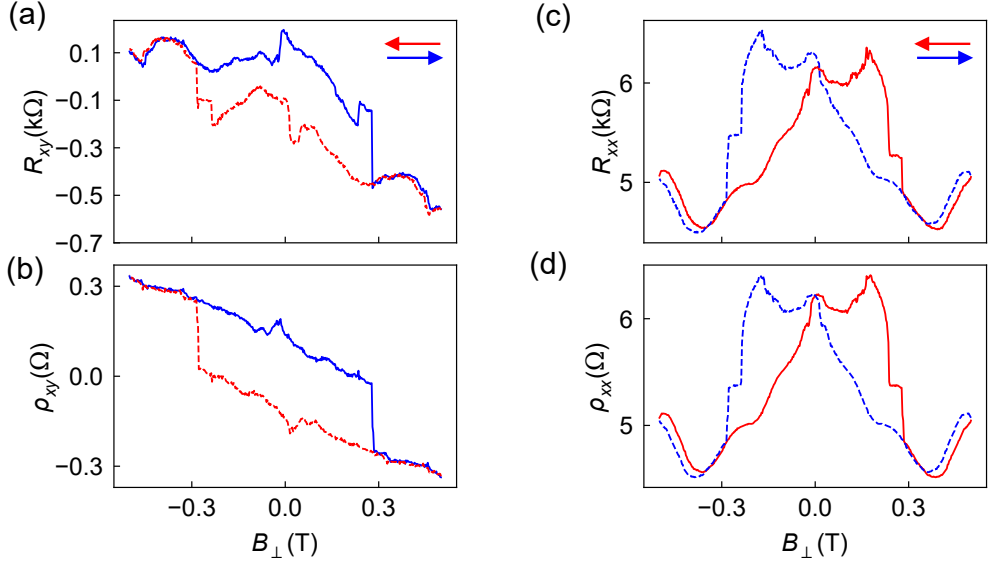

FIG. S8. **Symmetrization of $R_{xx}$ and Anti-symmetrization of $R_{xy}$.** (a) Measured Hall resistance $R_{xy}$ as a function of $B_\perp$, where the magnetic field is swept from $+B_\perp$ to $-B_\perp$ (red solid curve) and $-B_\perp$ to $+B_\perp$ (blue dashed curve). The corresponding anti-symmetrized Hall Resistance $\rho_{xy}$ is shown in (b). (c) Shows the measured longitudinal resistance $R_{xx}$ and the symmetrized longitudinal resistance, $\rho_{xx}$ in (d). These data were taken at $\nu = 3.67$ and $D = -0.43 \text{B/nm}$.

## IX. Anomalous Hall effect for positive displacement field, $+D$, device D1

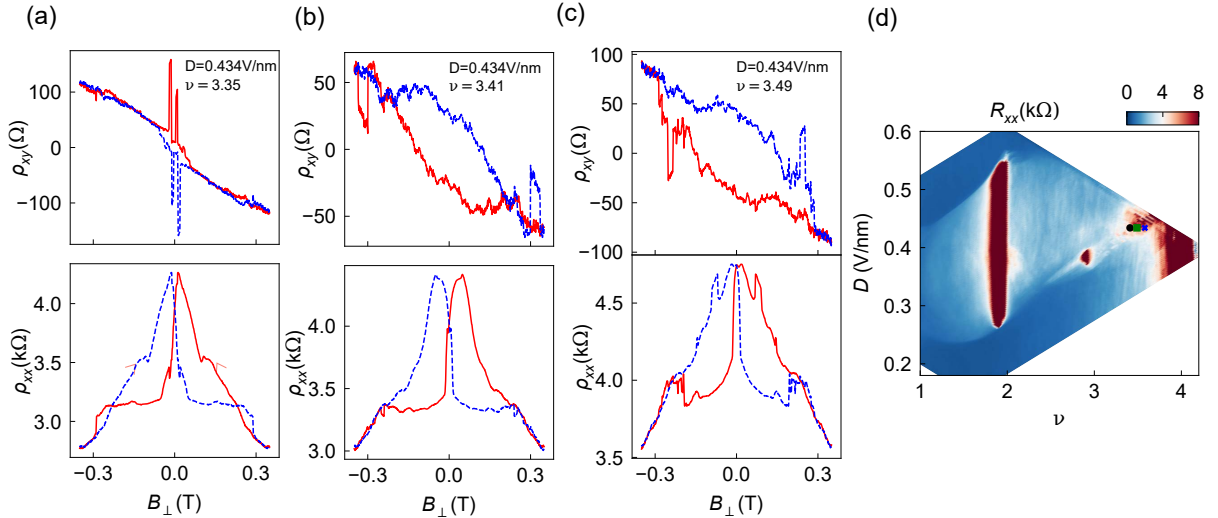

FIG. S9. **Anomalous Hall effect for positive displacement field $(+D)$.** Due to pinching off of voltage probe contacts it was difficult to access the AHE part of the $+D$ halo region. As shown in Fig. S9d, the AHE corner is just at the edge of the measurable region (contacts pinched off outside this). Even in the limited range that was available, AHE was clearly observed. (a) - (c) Anti-symmetrized Hall resistance and symmetrized longitudinal resistances for for several values of $\nu$ and $+D$, with corresponding locations shown in (d).

# X. Additional data: magnetic anisotropy for D1 and D2

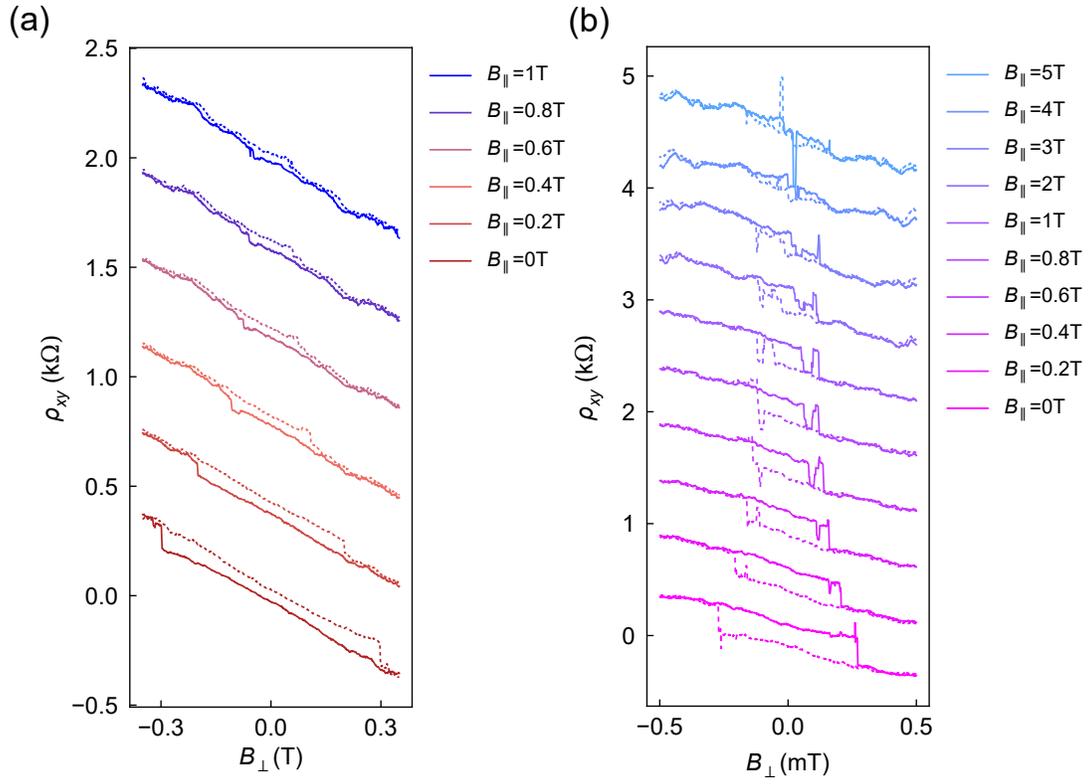

FIG. S10. **Magnetic anisotropy compared for samples D1 and D2.** (a) Out of plane hysteresis for several values of $B_{\parallel}$ for filling $\nu = 3.47$ and $D = -0.43$V/nm in device D1. (b) Out of plane hysteresis for several values of $B_{\parallel}$ at filling $\nu = 3.57$ and $D = -0.43$V/nm in device D2. This data is acquired for different pair of contacts as shown in Fig.3a of the main manuscript.

## XI. Temperature dependence of the correlated insulating states, device D1

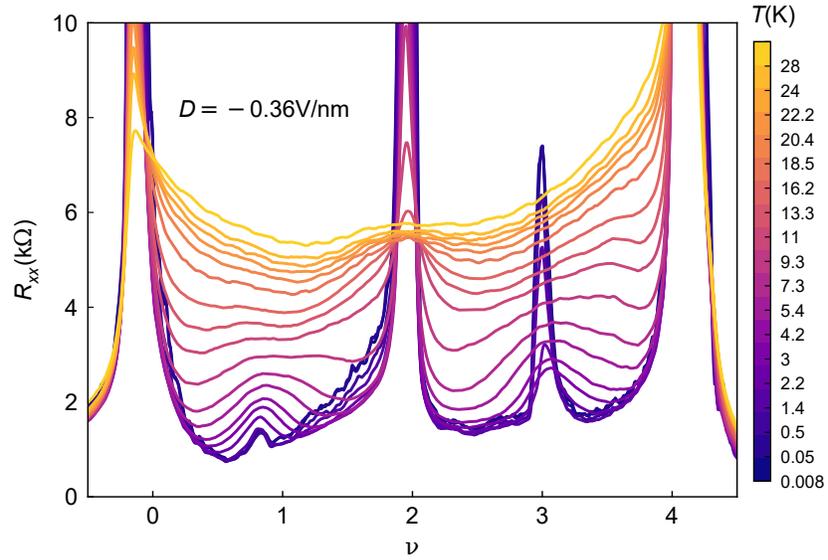

FIG. S11. **Temperature dependence.** Temperature dependence of the correlated insulating states, device D1. The correlated insulating state at $\nu = 2$ appears for $T < 15$K, while the correlated insulating states for $\nu = 3$ appears for $T < 2$K.

## XII. Collapse of AHE with temperature, device D1

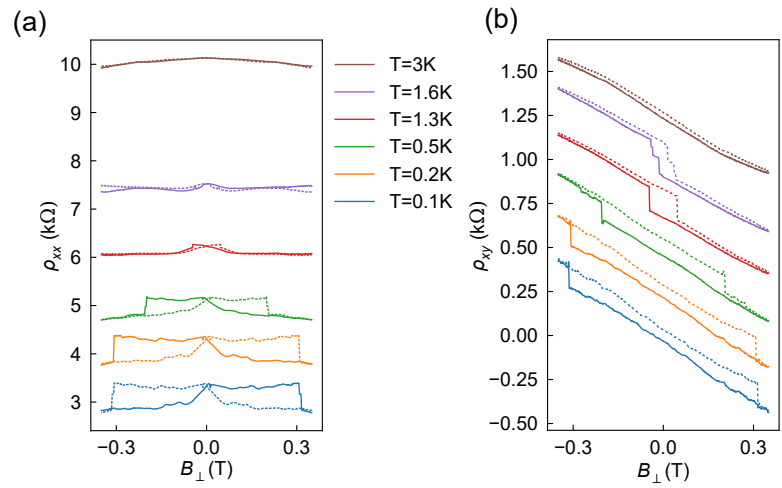

FIG. S12. **Temperature dependence of AHE, D1.** Temperature dependence of the hysteresis loop in $\rho_{xy}$ for $\nu = 3.58$, and $D = 0.42$ V/nm. Here each curves are shifted by $250\Omega$ for clarity. (b) Shows the temperature dependence of $\rho_{xx}$, where each curves are offset by $1\text{k}\Omega$ for clarity.

## XIII. Fig 2, 3, 4 data without B correction

The magnetic field axis in Figs 2 and 3 is rescaled to account for trapped flux in our superconducting magnet. The rescaling function maps $B_{real}$, the actual magnetic field estimated at the sample and the one plotted on the horizontal axes in the main text, to $B_{mag}$, the field set by the current in the magnet. The mapping depends on the sweep range and sweep direction, as the amount of trapped flux is greater for larger sweeps. The rescaling functions were estimated by examining $R_{xy}$ during magnetic field sweeps outside of the halo regions.

Scaling functions are given below, expressing $B$ in mT:

- For ±500 mT scans: $B_{real} = B_{mag} \pm (12.5 - 12.5 * (B_{mag}/500)^2)$
- For ±350 mT scans: $B_{real} = B_{mag} \pm (11 - 11 * (B_{mag}/350)^2)$
- For ±200 mT scans: $B_{real} = B_{mag} \pm (8.5 - 8.5 * (B_{mag}/200)^2)$
- In all cases, the + in the ± corresponds to downsweeps and the − corresponds to upsweeps.
- We estimate a remaining inaccuracy less than ±2 mT in the actual value of $B_{real}$ compared to $B_{mag}$.

For clarity, the raw (unshifted) data for Figs 2, 3 and 4 are shown below.

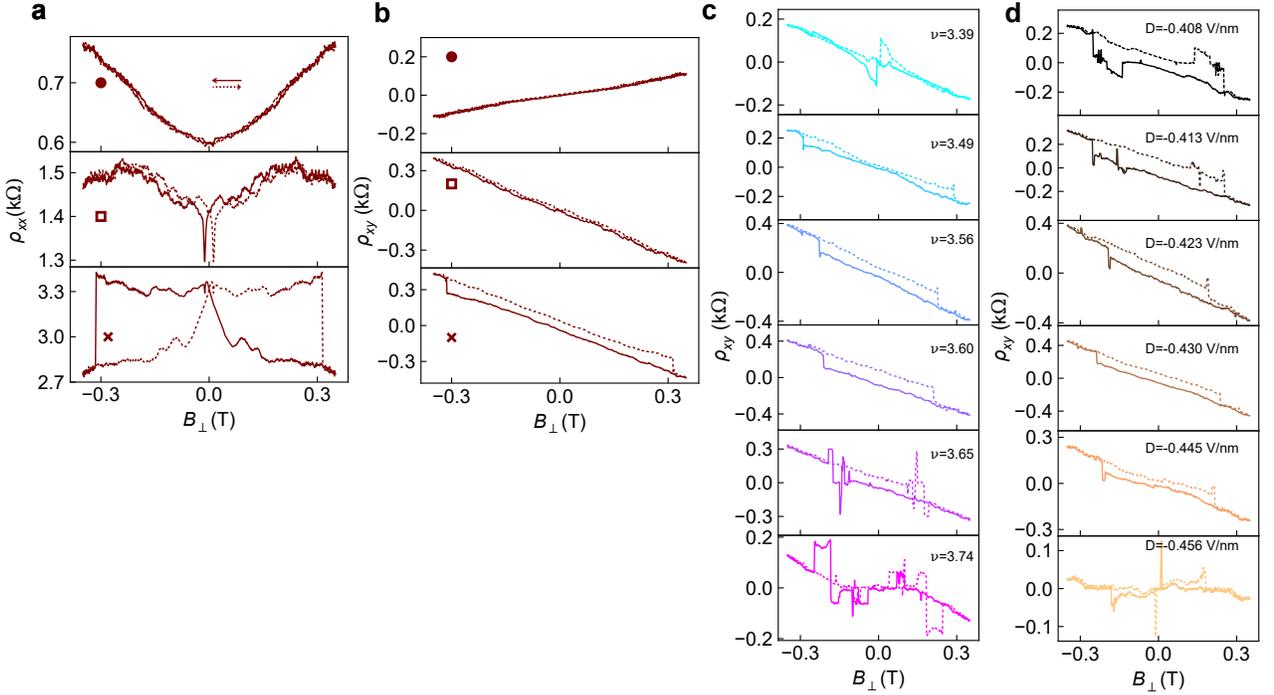

FIG. S13. **Fig 2 without magnetic field correction.** (a)-(d) Data of Fig 2 plotted against $B_{mag}$, that is, without magnetic field correction.

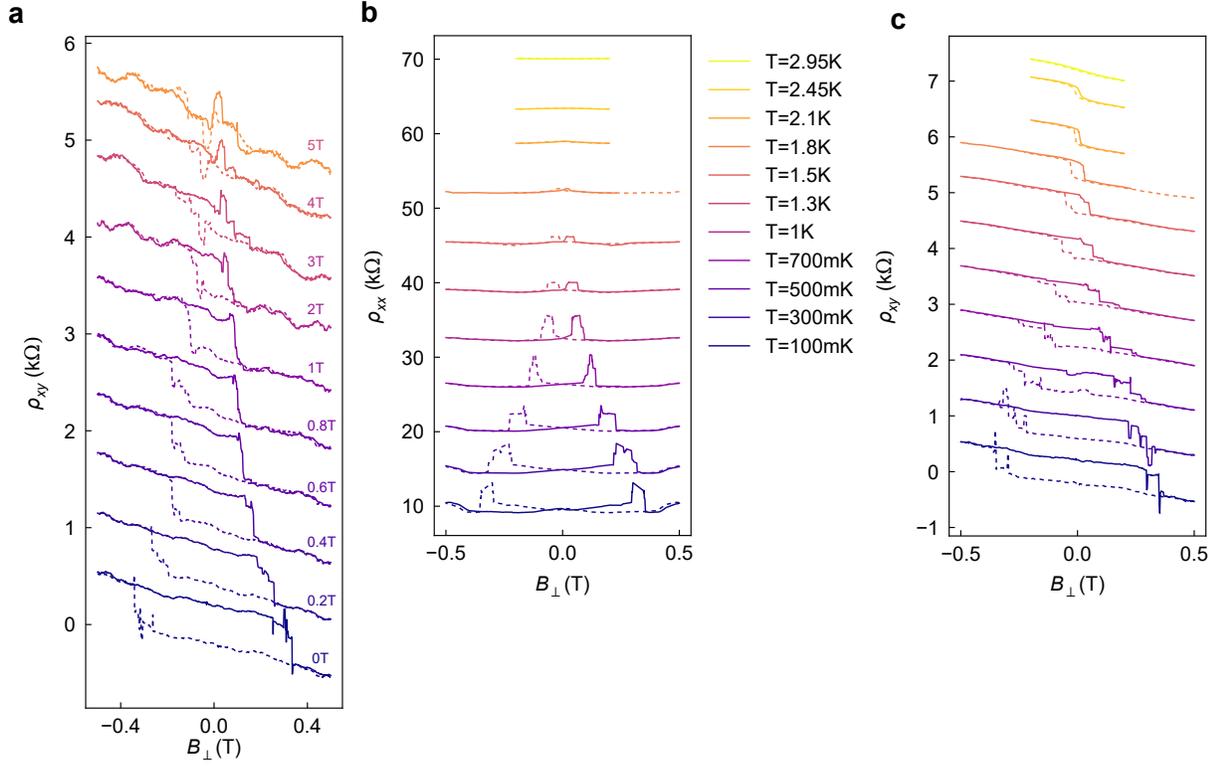

FIG. S14. **Fig 3 and 4 without magnetic field correction.** (a) Data of Fig. 3a plotted against $B_{mag}$, that is, without magnetic field correction. (b)-(c) Fig 4a and 4b of the main manuscript without magnetic field correction.

# SUPPLEMENTARY REFERENCES



\end{document}